\documentclass[
aps,
prc,
showpacs,
twocolumn,
superscriptaddress,
nofootinbib,
floatfix]
{revtex4-1}
\usepackage{graphicx,
longtable,
slashed,
hyperref,
bm,url,
braket,
xcolor,
xspace
}

\usepackage{bm}

\def\be{\begin{equation}}
\def\ee{\end{equation}}
\def\dg{\dagger}
\def\bee{\begin{eqnarray}}
\def\eee{\end{eqnarray}}

\newcommand{\lf}{\left(}
\newcommand{\rh}{\right)}

 \newcommand{\gd}[1]{\gamma_{#1}}

\begin{document}

\title{The $\lambda$ mechanism of the $0\nu\beta\beta$-decay }

\author{Fedor \v{S}imkovic}
\affiliation{Department of Nuclear Physics and Biophysics, Comenius
University, Mlynsk\'{a} dolina F1, SK-842 48
Bratislava, Slovakia}
\affiliation{Boboliubov Laboratory of Theoretical Physics, JINR 141980 Dubna,
Russia}
\affiliation{Czech Technical University in Prague, 128-00 Prague, Czech Republic}
\author{Du\v{s}an \v Stef\'{a}nik}
\affiliation{Department of Nuclear Physics and Biophysics, Comenius
University, Mlynsk\'{a} dolina F1, SK-842 48
Bratislava, Slovakia}
\author{Rastislav Dvornick\'{y}}
\affiliation{Department of Nuclear Physics and Biophysics, Comenius
University, Mlynsk\'{a} dolina F1, SK-842 48
Bratislava, Slovakia}
\affiliation{Dzhelepov Laboratory of Nuclear Problems, JINR 141980 Dubna,
Russia}

\begin{abstract}
The $\lambda$ mechanism ($W_L$-$W_R$ exchange) of the neutrinoless double beta
decay ($0\nu\beta\beta$-decay), which has 
origin in left-right symmetric model with right-handed gauge boson at TeV scale,
is investigated. The revisited formalism of the $0\nu\beta\beta$-decay, which 
includes higher order terms of nucleon current, is exploited. The
corresponding nuclear matrix elements are
calculated within quasiparticle random phase approximation with partial
restoration of the isospin symmetry for nuclei of experimental interest.
A possibility to distinguish between the conventional light neutrino
mass ($W_L$-$W_L$ exchange) and $\lambda$ mechanisms by observation of
the $0\nu\beta\beta$-decay in several nuclei is discussed. 
A qualitative comparison of effective lepton number violating couplings
associated with these two mechanisms is performed. By making viable assumption
about the seesaw type mixing of light and heavy neutrinos with 
the value of Dirac mass $m_D$ within the range $1~ \textrm{MeV} < m_D < 1~ \textrm{GeV}$,
it is concluded that there is a dominance of the conventional
light neutrino mass mechanism in the decay rate.
\end{abstract}
\medskip

\maketitle

%
\section{Introduction}
%

The Majorana nature of neutrinos, as favored by many theoretical models, is a key for understanding
of tiny neutrino masses observed in neutrino oscillation experiments. A golden process for answering
this open question of particle physics is the neutrinoless double beta decay
($0\nu\beta\beta$-decay) \cite{review,vissa16,ves16},
\begin{equation}
(A,Z) \rightarrow (A,Z+2) + 2 e^-, 
\end{equation}
in which an atomic nucleus with Z protons decays to another one with two more protons and
the same mass number A, by emitting two electrons and nothing else. The observation of this
process, which violates total lepton number conservation and is forbidden in the Standard Model,
guaranties that neutrinos are Majorana particles, i.e., their own antiparticles \cite{schechter}.

The searches for the $0\nu\beta\beta$-decay have not yielded any evidence for Majorana neutrinos yet.
This could be because neutrinos are Dirac particles, i.e. not their own antiparticles.
In this case we will never observe the decay. However, it is assumed that the reason for it is 
not sufficient sensitivity of previous and current $0\nu\beta\beta$-decay experiments to
the occurrence of this rare process. 

Due to the evidence for neutrino oscillations and therefore for 3 neutrino mixing and masses 
the $0\nu\beta\beta$-decay mechanism of primary interest is the exchange of 3 light Majorana neutrinos
interacting through the left-handed V-A weak currents ($m_{\beta\beta}$ mechanism).
In this case, the inverse $0\nu\beta\beta$-decay
half-life is given by \cite{review,vissa16,ves16}
\be \label{facbez}
\left[T_{1/2}^{0\nu} \right]^{-1} = \left(\frac{m_{\beta\beta}}{m_e}\right)^2~ 
g_A^4 ~M_\nu^2 ~G_{01},
\ee
where $G_{01}$, $g_A$ and $M_\nu$ represent  an exactly calculable phase space factor, 
the axial-vector coupling constant and the nuclear matrix element (whose calculation
represents a severe challenge for nuclear theorists), respectively. $m_e$ is the mass of an electron.
The effective neutrino mass,
\begin{equation}
m_{\beta\beta} = \left|U^2_{e 1} m_1 + U^2_{e 2} m_2 + U^2_{e 3} m_3\right|,
\label{mbb}
\end{equation}
is a linear combination of the three neutrino masses $m_i$, weighted with the square of the elements
$U_{e i}$ of the first row of the Pontecorvo-Maki-Nakagawa-Sakata (PMNS) neutrino mixing matrix. The
measured value of $m_{\beta\beta}$ would be a source of important information about
the neutrino mass spectrum (normal or inverted spectrum), absolute neutrino mass scale and 
the CP violation in the neutrino sector. However, that is not the only possibility.

There are several different theoretical frameworks that provide various $0\nu\beta\beta$-decay
mechanisms, which generate masses of light Majorana neutrinos and violate the total lepton number
conservation.
One of those theories is the left-right symmetric model (LRSM) \cite{Pati,Mohap},
in which corresponding to the left-handed neutrino, there is a parity symmetric right-handed neutrino.
The parity between left and right is restored at high energies and neutrinos acquire mass
through the see-saw mechanism, what requires presence of additional heavy neutrinos. 
In general one cannot predict the scale where the left-right symmetry is realized, which
might be as low as a few TeV - accessible at Large Hadron Collider, or as large
as GUT scale of $10^{15}$ GeV. 

The LRSM, one of the most elegant theories beyond the Standard Model, offers a number
of new physics contributions to $0\nu\beta\beta$-decay, either from right-handed neutrinos
or Higgs triplets. The main question is whether these additional $0\nu\beta\beta$-mechanisms
can compete with the $m_{\beta\beta}$ mechanism and affect the $0\nu\beta\beta$-decay  
rate significantly. This issue is a subject of intense theoretical investigation
within the TeV-scale left-right symmetry theories \cite{Tello,Barry13,Dev,deppisch,borah}.
In analysis of heavy neutrino mass mechanisms of the $0\nu\beta\beta$-decay  
an important role plays a study of related lepton number and lepton flavor
violation processes in experiments at Large Hadron Collider
\cite{vissa16,nemev11,helo13,gluza15,gonz16,gluza16}.

The goal of this article is to discuss in details the $W_L-W_R$ exchange mechanism of
the $0\nu\beta\beta$-decay mediated by light neutrinos ($\lambda$ mechanism) and its
coexistence with the standard $m_{\beta\beta}$ mechanism. For that purpose the corresponding
nuclear matrix elements (NMEs) will be calculated within the quasiparticle random
phase approximation with a partial restoration of the isospin symmetry \cite{vadimp}
by taking the advantage of improved formalism for this mechanism of the
$0\nu\beta\beta$-decay of Ref. \cite{StefA}. A possibility to distinguish $m_{\beta\beta}$
and $\lambda$  mechanisms in the case of observation of the $0\nu\beta\beta$-decay on several isotopes
will be analyzed. 
Further, the dominance of any of these two mechanisms in the $0\nu\beta\beta$-decay
rate will be studied within seesaw model with right-handed gauge boson at TeV scale.
We note that a similar analysis was performed by exploiting
a simplified $0\nu\beta\beta$-decay rate formula and different viable particle physics scenarios
in Refs. \cite{Tello,Barry13,Dev,deppisch,borah}.

%
\section{Decay rate for the neutrinoless double-beta decay}
%

Recently, the $0\nu\beta\beta$-decay with the inclusion of right-handed leptonic and
hadronic currents has been revisited by considering exact Dirac wave function with
finite nuclear size and electron screening of emitted electrons and the induced pseudoscalar
term of hadron current, resulting in additional nuclear matrix elements \cite{StefA}.
In this section we present the main elements of the revisited formalism of the
$\lambda$ mechanism of the $0\nu\beta\beta$-decay briefly. Unlike in \cite{StefA}
the effect the weak-magnetism term of the hadron current on leading NMEs
is taken into account. 

If the mixing between left and right vector bosons is neglected,
for the effective weak interaction hamiltonian density generated within
the LRSM we obtain 
\bee \label{hamilweakrh}
H^\beta &=& \frac{G_{\beta}}{\sqrt{2}} ~\left[
  j_L^{~\rho}J^{\dg}_{L\rho } +\lambda j_R^{~\rho}J^{\dg}_{R\rho } +  h.c.\right].
\eee
Here, $G_{\beta}=G_F\cos{\theta_C}$, where $G_F$ and $\theta_C$ are Fermi constant and Cabbibo angle, respectively.
The coupling constant $\lambda$ is defined as
\bee \label{relacieLRSM}
\lambda &=&  (M_{W_L}/M_{W_R})^2.
\eee
Here, $M_{W_L}$ and $M_{W_R}$ are masses of the Standard Model left-handed $W_L$ 
and right-handed $W_R$ gauge bosons, respectively.
The left- and right-handed leptonic currents are given by 
\bee
j_L^{~\rho}=\bar{e}\gd{\rho}(1-\gd{5})\nu_{eL}, \qquad j_R^{~\rho}=\bar{e}\gd{\rho}(1+\gd{5})\nu_{eR}.\nonumber \\
\eee
The weak eigenstate electron neutrinos $\nu_{eL}$ and  $\nu_{eR}$ are superpositions of the light and heavy 
 mass eigenstate Majorana neutrinos $\nu_j$ and $N_j$, respectively. We have
 \begin{eqnarray}\label{lambdaeta}
\nu_{eL} &=& \sum_{j=1}^3\lf U_{ej}\nu_{jL}+S_{ej} (N_{jR})^C \rh,\nonumber\\
\nu_{eR} &=& \sum_{j=1}^3 \lf T_{ej}^*(\nu_{jL})^C+V_{ej}^*N_{jR} \rh.
\end{eqnarray}
Here, $U, S, T,$ and  $V$ are the $3\times 3$ block matrices in  flavor  space, which 
constitute a generalization of the Pontecorvo-Maki-Nakagawa-Sakata matrix, namely
the $6\times 6$ unitary neutrino mixing matrix \cite{Xing}
\bee
\mathcal{U}&=& \left(
\begin{array}{ll}
U & S\\
 T &V \\
 \end{array}
\right). 
\label{maticazmies}
\eee
The nuclear currents are, in the non-relativistic approximation,
\cite{Doies}
\begin{eqnarray}
\label{nonhad}
J^{\rho}_{L,R}(\mathbf{x}) &=& \sum_{n}\tau_{n}^+\delta(\mathbf{x}-\mathbf{r}_n)
\left[\left(g_{V}\mp g_{A}C_n\right)g^{\rho 0} \right.\nonumber\\
&&~\left.  + g^{\rho k}\left(\pm g_{A}\sigma_n^k-g_{V}D_n^k
  \mp g_{P}~q_n^{k}~\frac{\vec{\sigma}_n \cdot \mathbf{q}_n}{2m_N}
  \right)\right].\nonumber\\
\end{eqnarray}
Here,  $m_N$ is the nucleon mass. $q_V\equiv q_V(q^2)$, $q_A\equiv q_A(q^2)$, $q_M\equiv q_M(q^2)$ and $q_P\equiv q_P(q^2)$
are, respectively, the vector, axial-vector, weak-magnetism and induced pseudoscalar form-factors. 
The nucleon recoil terms are given by
\begin{eqnarray}
C_n &=& \frac{\vec{\sigma} \cdot \left(\mathbf{p}_n+\mathbf{p}_n^{'}\right)}{2m_N}-\frac{g_P}{g_A}\left(E_n-E_n^{'}\right)
\frac{\vec{\sigma} \cdot \mathbf{q}_n}{2m_N},\nonumber\\
\mathbf{D}_n &=& \frac{\left(\mathbf{p}_n+\mathbf{p}_n^{'}\right)}{2m_N}-i\left(1+\frac{g_M}{g_V}\right)
\frac{\vec{\sigma} \times \mathbf{q}_n}{2m_N},
\end{eqnarray}
where $\mathbf{q}_n=\mathbf{p}_n-\mathbf{p}_n^{'}$ is the momentum transfer between the nucleons.
The initial neutron (final proton) possesses  energy $E_n'$ ($E_n$)
and momentum $\mathbf{p}_n'$ ($\mathbf{p}_n$). $\mathbf{r}_n$, $\tau^+_n$ and  $\vec{\sigma}_n$,
which act on the $n$-th nucleon, are the position operator, the isospin raising operator  and the Pauli matrix,  
respectively. 

By assuming standard approximations \cite{StefA} for the $0\nu\beta\beta$-decay half-life we get
\begin{eqnarray} \label{halflifem}
&&\left[T_{1/2}^{0\nu}\right]^{-1} = \eta_\nu^2 ~C_{mm} + \eta_\lambda^2~ C_{\lambda\lambda}
+ \eta_\nu ~\eta_\lambda ~\cos{\psi} ~C_{m\lambda}.
\end{eqnarray}
The effective lepton number violating parameters $\eta_\nu$ ($W_L-W_L$ exchange),
$\eta_\lambda$ ($W_L-W_R$ exchange) and their relative phase $\Psi$ are given by
\begin{eqnarray}\label{effective}
\eta_\nu &=& \frac{m_{\beta\beta}}{m_e},~~~
\eta_\lambda = \lambda|\sum_{j=1}^3 U_{ej}T_{ej}^*|,\nonumber\\
\psi  &=& \textrm{arg}[(\sum_{j=1}^3 m_j U_{ej}^2 )(\sum_{j=1}^3 U_{ej}T_{ej}^*)^{*}].
\end{eqnarray}
The coefficients $C_I$ (I = $mm$, $m\lambda$ and $\lambda\lambda$)
are linear combinations of products of nuclear matrix elements and phase-space factors:
\bee\label{Ccoef}
C_{mm} &=& g_A^4 M_\nu^2 ~G_{01}, \nonumber\\
C_{m\lambda} &=& - g_A^4 M_\nu \left( M_{2-} G_{03} -  M_{1+} G_{04}\right), \nonumber\\
C_{\lambda\lambda} &=& g_A^4 \left( M_{2-}^2 G_{02}
+ \frac{1}{9} M_{1+}^2 G_{011} - \frac{2}{9} M_{1+} M_{2-} G_{010} \right),\nonumber\\
\eee
The explicit form and calculated values of phase-space factors
$G_{0i}$ (i=1, 2, 3, 4, 10 and 11) of the
$0\nu\beta\beta$-decaying nuclei of experimental interest are given in \cite{StefA}.
The NMES, which constitute the coefficients $C_I$ in Eq. (\ref{Ccoef}),
are defined as follows:
\bee
M_\nu  &=&  M_{GT} - \frac{M_{F}}{g_A^2} + M_{T}, \nonumber\\
M_{\nu\omega} &=& M_{GT\omega} - \frac{M_{F\omega}}{g_A^2} + M_{T\omega} \nonumber\\
M_{1+ } &=& M_{qGT} + 3 \frac{M_{qF}}{g_A^2} - 6 M_{qT}, \nonumber\\
M_{2- } &=& M_{\nu\omega} - \frac{1}{9} M_{1+}.
\label{nulamnme}
\eee
The partial nuclear matrix elements $M_I$, where
$\textrm{I=GT, F, T, $\omega$F, $\omega$GT, $\omega$T, qF, qGT, and qT}$
are given by 
\begin{eqnarray}
M_{ F,  GT,  T} &=& \sum_{rs}\langle A_f| h_{  F,   GT, T}(r_-) {O }_{F,GT,T} |A_i\rangle\nonumber\\
M_{\omega F, \omega GT, \omega T} &=& \sum_{rs} \langle A_f | h_{ \omega F,  \omega GT,\omega T}(r_-) {O}_{F,GT,T} | A_i\rangle\nonumber\\
M_{qF,qGT,qT} &=& \sum_{rs} \langle A_f | h_{qF,qGT,qT}(r_-){O}_{F,GT,T} | A_i\rangle. 
\end{eqnarray}
Here, ${O}_{F,GT,T}$ are the Fermi, Gamow-Teller and tensor operators $1, \vec{\sigma}_1 \cdot \vec{\sigma}_2$ and  
$3(\vec{\sigma}_1 \cdot \hat{r}_{12})(\vec{\sigma}_2 \cdot \hat{r}_{12})$.
The two-nucleon exchange potentials $h_I(r)$ with  =F, GT, T, $\omega$F,  $\omega$GT,  $\omega$T,
qF, qGT, and qT can be written as
\bee
&&h_I(r) = \frac{2R}{\pi}\int f_I(q,r) \frac{q ~dq}{q +\bar{E}_n -(E_i+E_f)/2}, 
\eee
where
\bee
f_{GT} &=&  \frac{j_0(q,r)}{g^2_A}
\left(g_A^2(q^2) - \frac{g_A(q^2)g_P(q^2)}{m_N}\frac{q^2}{3} \right. \nonumber\\
&&~~\left. +\frac{g_P^2(q^2)}{4m^2_N}\frac{q^4}{3} 
+ 2 \frac{g_M^2(q^2)}{4m^2_N}\frac{q^2}{3}\right),\nonumber\\
f_{F} &=& g_V^2(q^2)j_0(q r), \nonumber\\
f_{T} &=&  \frac{j_2(q,r)}{g^2_A}
\left(\frac{g_A(q^2)g_P(q^2)}{m_N}\frac{q^2}{3}-\frac{g_P^2(q^2)}{4m^2_N}\frac{q^4}{3} \right.\nonumber\\
&&~~ \left. + \frac{g_M^2(q^2)}{4m^2_N}\frac{q^2}{3}\right),\nonumber\\
f_{qF}&=&r g_V^2(q^2)j_1(qr)q,\nonumber \\
f_{qGT}&=&\left(\frac{g_A^2(q^2)}{g_A^2} q+3\frac{g_P^2(q^2)}{g_A^2}\frac{q^5}{4m_N^2}\right. \nonumber\\
&& ~~ \left. + \frac{g_A(q^2)g_P(q^2)}{g_A^2}\frac{q^3}{m_N}\right)r j_1(qr),\nonumber\\
f_{qT}&=& \frac{r}{3}\left(\left(\frac{g_A^2(q^2)}{g_A^2} 
q-\frac{g_P(q^2)g_A(q^2)}{2g_A^2}\frac{q^3}{m_N}\right)j_1(qr)\right.\nonumber \\
&-&\left.9\frac{g_P^2(q^2)}{2g_A^2}\frac{ q^5}{20m_N^2}\left[ 2j_1(qr)/3-j_3(qr)\right]\right),
\eee
and
\bee
h_{\omega GT, F, T}= \frac{q}{\left(q +\bar{E}_n -(E_i+E_f)/2\right)} h_{GT, F, T}.
\eee
Here, $E_i$, $E_f$  and $\bar{E}_n$ are energies of the initial and final nucleus and averaged energy of
intermediate nuclear states, respectively.
$\mathbf{r}= (\mathbf{r}_r - \mathbf{r}_s)$,  $\mathbf{r}_{r,s}$ is the coordinate of decaying nucleon and 
$j_i(qr)$ (i=1,2,3) denote the spherical Bessel functions. 
$\mathbf{p}_r+\mathbf{p}_r' \simeq  0$, $E_r-E_r' \simeq  0$ and $\mathbf{p}_r-\mathbf{p}_r' \simeq \mathbf{q}$,
where $\mathbf{q}$ is the momentum exchange. The form factors $g_V(q^2)$, $g_A(q^2)$, $g_M(q^2)$ and $g_P(q^2)$ are defined in
\cite{Simdva}. We note that factor 4  in definition of the two-nucleon exchange potentials $h_I(r)$ with
I=$\omega$F, $\omega$GT, and $\omega$T in Eq. (48) of Ref. \cite{StefA} needs to be replaced by factor 2.

\begin{center}
  \begin{figure}[htb]
  \includegraphics[width=8cm]{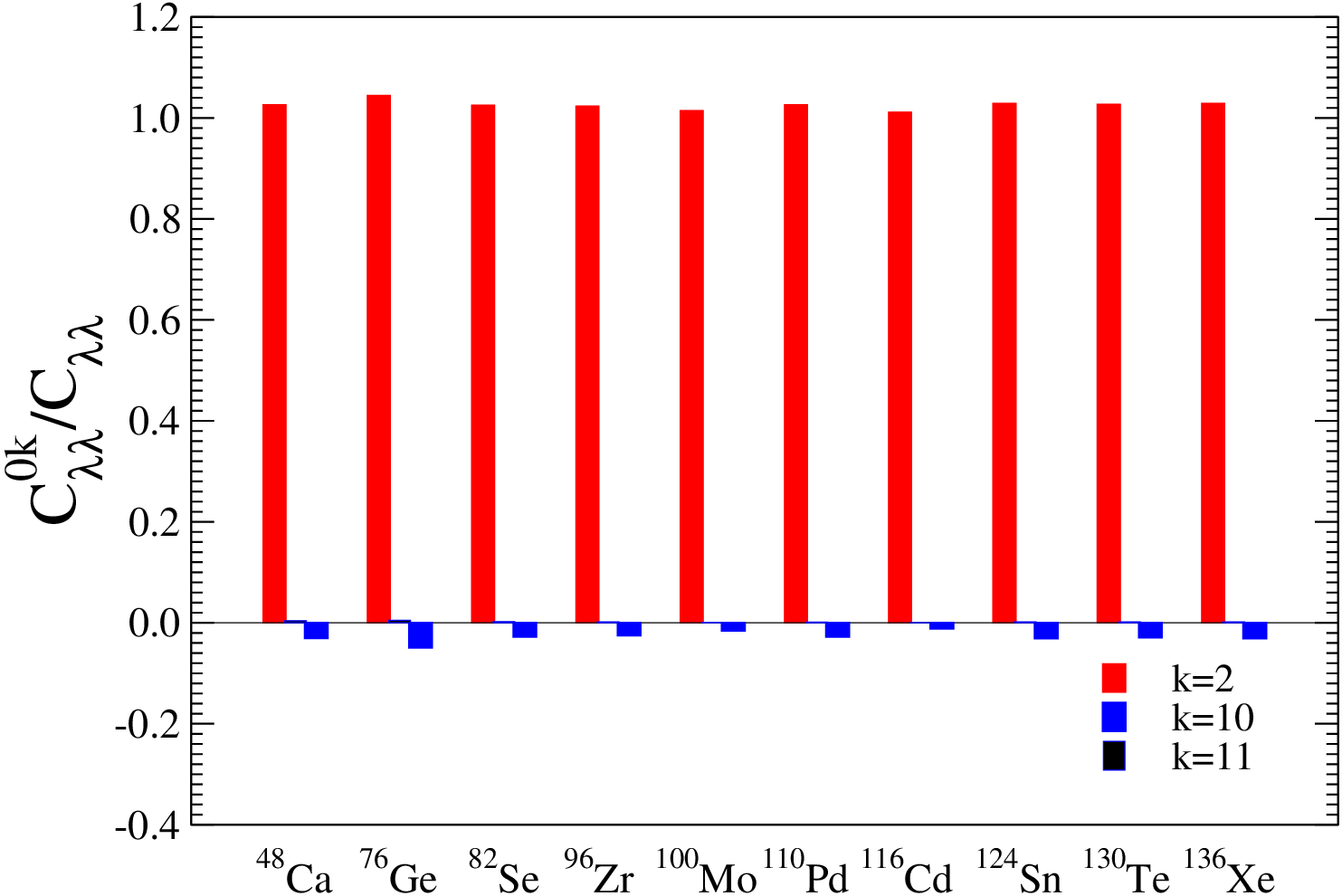}    
    \caption{(Color online) The decomposition of coefficient 
    $C_{\lambda\lambda}$ (see Eq. (\ref{halflifem})) on partial contributions $C^{0k}_I$ associated with
    phase-space factors $G_{0k}$ (k=2, 10 and 11) for nuclei of experimental interest.
    The partial contributions are identified by index k. The contributions from largest to
    the smallest are  displayed in red, blue and black colors, respectively. 
  }
\label{fig:Ccomp}
  \end{figure}
\end{center}

\begin{widetext}

\begin{table}[!t]
  \begin{center}
    \caption{The nuclear matrix elements of the $0\nu\beta\beta$-decay associated with
      $m_{\beta\beta}$  and $\lambda$  mechanisms and the coefficients
      $C_{mm}$, $C_{m\lambda}$ and $C_{\lambda\lambda}$  (in $10^{-14}$ yrs$^{-1}$) of the
      decay rate formula (see Eq. (\ref{halflifem})). The nuclear matrix elements are 
      calculated within the quasiparticle random phase approximation with partial restoration
      of the isospin symmetry. The G-matrix elements of a realistic Argonne V18 nucleon-nucleon 
      potential are considered \cite{vadimp}. The phase-space factors  are taken from 
      \cite{StefA}. $f_{\lambda m}= C_{\lambda\lambda}/C_{mm}$, $f^G_{\lambda m}= G_{02}/G_{01}$ and $g_A=1.269$
      is assumed. $Q_{\beta\beta}$ is the Q-value of the double beta decay in MeV.
      \label{tab.MatElem}}
 \begin{tabular}{lccccccccccc}\hline\hline
 & & $^{48}$Ca & $^{76}$Ge & $^{82}$Se & $^{96}$Zr & $^{100}$Mo & $^{110}$Pd & $^{116}$Cd & $^{124}$Sn & $^{130}$Te &
   $^{136}$Xe   \\ \hline
                    & &     &     &     &     &      &     &       &      &     &   \\            
   & &    \multicolumn{10}{c}{pnQRPA NMEs of Ref. \cite{muto89}}    \\
                    & &     &     &     &     &      &     &       &      &     &   \\            
   $M_{GT}$         & &
      &    3.014  &     2.847 &           &     0.763 & 
      &           &           &     2.493 &     1.120 \\   
   $M_{F}$         & &
      &    -1.173 &    -1.071 &           &    -1.356 & 
      &           &           &    -0.977 &    -0.461 \\    
                    & &     &     &     &     &      &     &       &      &     &   \\         
   $M_{\omega GT}$         & &
      &     2.912 &     2.744 &           &     1.330 & 
      &           &           &     2.442 &     1.172 \\    
   $M_{\omega F}$         & &
      &    -1.025 &    -0.939 &           &    -1.218 & 
      &           &           &    -0.867 &    -0.411 \\    
                    & &     &     &     &     &      &     &       &      &     &   \\         
   $M_{q GT}$         & &
      &     1.945 &     1.886 &           &    -1.145 & 
      &           &           &     1.526 &     0.480 \\    
   $M_{q F}$         & &
      &    -1.058 &    -0.966 &           &    -1.161 & 
      &           &           &    -0.860 &    -0.389 \\
                    & &     &     &     &     &      &     &       &      &     &   \\            
   & &    \multicolumn{10}{c}{Present work}    \\
                    & &     &     &     &     &      &     &       &      &     &   \\            
   $M_{GT}$         & &
      0.569 &     4.513 &     4.005 &     2.104 &     4.293 & 
      4.670 &     3.178 &     2.056 &     3.192 &     1.808 \\   
   $M_{F}$         & &
     -0.312 &    -1.577 &    -1.496 &    -1.189 &    -2.214 & 
     -2.152 &    -1.573 &    -0.907 &    -1.489 &    -0.779 \\    
   $M_{T}$         & &
     -0.162 &    -0.571 &    -0.525 &    -0.397 &    -0.650 & 
     -0.558 &    -0.262 &    -0.350 &    -0.561 &    -0.288 \\    
                    & &     &     &     &     &      &     &       &      &     &   \\         
   $M_{\omega GT}$         & &
      0.568 &     4.238 &     3.784 &     2.088 &     4.159 & 
      4.436 &     2.979 &     2.108 &     3.091 &     1.758 \\    

   $M_{\omega F}$         & &
     -0.295 &    -1.487 &    -1.409 &    -1.117 &    -2.076 & 
     -2.015 &    -1.466 &    -0.955 &    -1.410 &    -0.745 \\    
   $M_{\omega T}$         & &
     -0.156 &    -0.547 &    -0.502 &    -0.379 &    -0.623 & 
     -0.535 &    -0.251 &    -0.368 &    -0.536 &    -0.275 \\    
                    & &     &     &     &     &      &     &       &      &     &   \\         
   $M_{q GT}$         & &
      0.245 &     2.919 &     2.533 &     1.026 &     2.389 & 
      2.878 &     2.105 &     1.109 &     1.746 &     0.975 \\    
   $M_{q F}$         & &
     -0.203 &    -1.071 &    -1.031 &    -0.804 &    -1.588 & 
     -1.565 &    -1.208 &    -0.617 &    -0.995 &    -0.492 \\    
   $M_{q T}$         & &
     -0.107 &    -0.294 &    -0.262 &    -0.200 &    -0.329 & 
     -0.281 &    -0.142 &    -0.156 &    -0.252 &    -0.125\\    
                    & &     &     &     &     &      &     &       &      &     &   \\      
   $M_\nu$           & &
      0.601 &     4.921 &     4.410 &     2.446 &     5.018 & 
      5.449 &     3.894 &     2.333 &     3.554 &     2.004 \\
   $M_{\nu\omega}$           & &
      0.595 &     4.615 &     4.157 &     2.402 &     4.826 & 
      5.153 &     3.638 &     2.269 &     3.430 &     1.946 \\
   $M_{1^+}$           & &
      0.506 &     2.689 &     2.183 &     0.729 &     1.402 & 
      1.646 &     0.705 &     0.894 &     1.407 &     0.807 \\   
   $M_{2^-}$           & &
      0.497 &     4.549 &     4.096 &     2.364 &     4.790 & 
      5.114 &     3.613 &     2.222 &     3.381 &     1.897 \\
                    & &     &     &     &     &      &     &       &      &     &   \\      
   $C_{mm}$        & &
      2.33 &    14.9 &    51.3 &    32.0 &   104. & 
      37.2 &    65.8 &    12.8 &    46.7 &   15.2 \\ 
   $C_{m\lambda}$       & &
    -1.04 &   -5.96 &   -27.0 &   -20.1 &   -62.2 & 
    -17.0 &   -38.1 &   -6.24 &   -24.0 &    -7.62 \\ 
   $C_{\lambda\lambda}$      & &
     10.1 &    20.1 &   150. &   128. &   339. & 
     53.8 &   179. &    24.5 &   109. &   33.4 \\
    & &     &     &     &     &      &     &       &      &     &   \\
    $Q_{\beta\beta}$   & &
 4.272 & 2.039  & 2.995  & 3.350    & 3.034   & 2.017  & 2.814 & 2.287 & 2.527  & 2.457  \\    
   $f_{\lambda m}$         & &
       4.344 &     1.349 &     2.917 &     4.002 &     3.256 & 
       1.446 &     2.723 &     1.913 &     2.324 &     2.191 \\
   $f^G_{\lambda m}$         & &
      6.536 &     1.650 &     3.467 &     4.345 &     3.628 & 
      1.685 &     3.197 &     2.171 &     2.639 &     2.516 \\
      \hline\hline
      \end{tabular}
  \end{center}
\end{table}

\end{widetext}      

\begin{center}
  \begin{figure}[!t]
  \includegraphics[width=8cm]{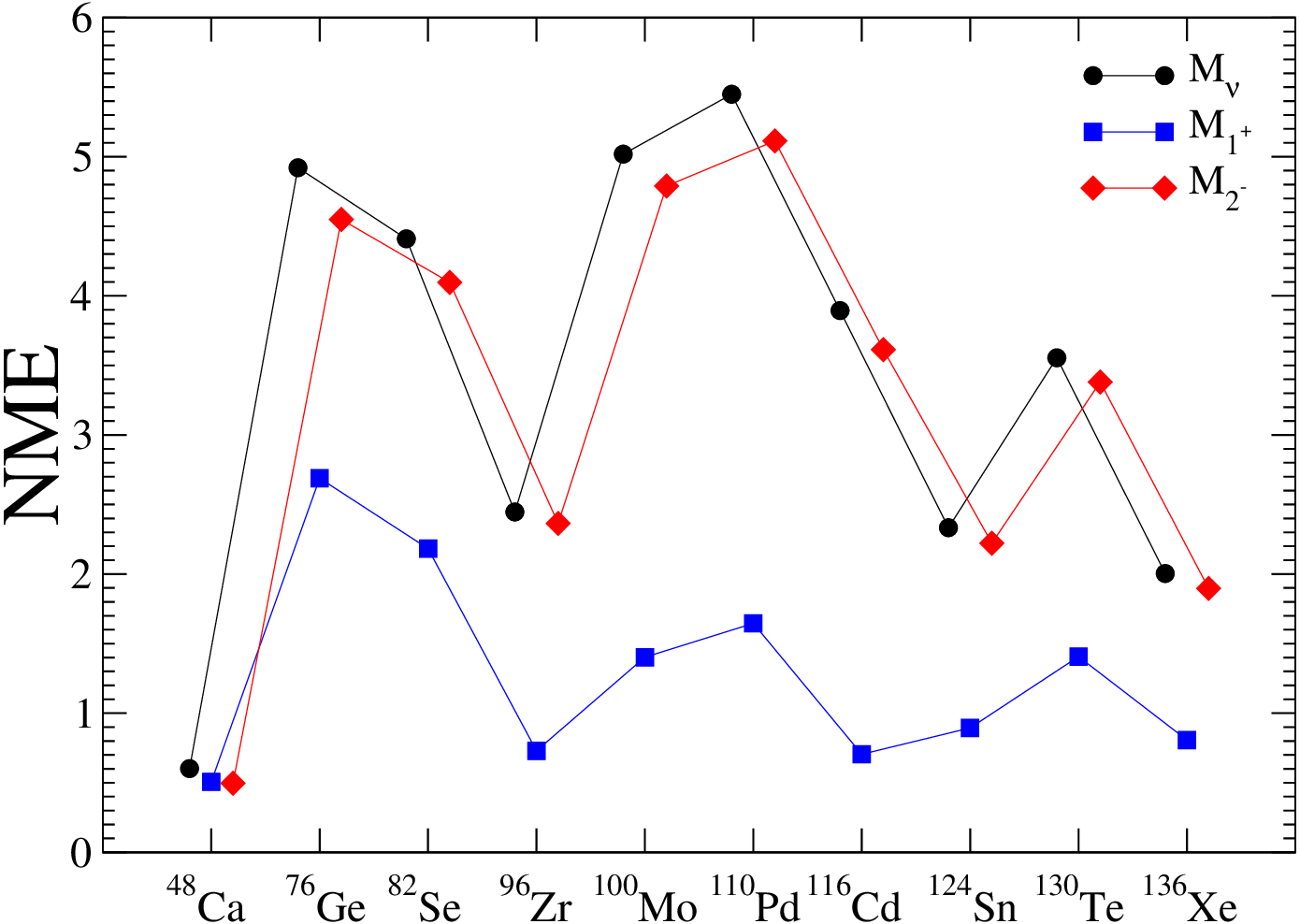}        
    \caption{(Color online) A comparison of the nuclear matrix elements
      $M_{1^+}$, $M_{2^-}$ ($\lambda$ mechanism) and $M_\nu$ ($m_{\beta\beta}$ mechanism)
      of the $0\nu\beta\beta$-decay. 
  }
\label{fig:nme}
\end{figure}
\end{center}

%
\section{Results and Discussion}
%

The nuclear matrix elements are calculated in proton-neutron quasiparticle 
random phase approximation with partial restoration of the isospin symmetry
for $^{48}$Ca, $^{76}$Ge, $^{82}$Se, $^{96}$Zr, $^{100}$Mo, $^{110}$Pd, $^{116}$Cd,
$^{124}$Sn, $^{130}$Te and $^{136}$Xe, which are of experimental interest.
In the calculation the same set of nuclear structure
parameters is used  as in \cite{vadimp}.
The pairing and residual interactions as well as the two-nucleon short-range
correlations derived from the realistic nucleon-nucleon Argonne V18 potential
are considered \cite{src09}. The closure approximation  for intermediate
nuclear states is assumed with $(\bar{E}_n -(E_i+E_f)/2) = 8$ MeV. 
The free nucleon value of axial-vector coupling constant  ($g_A=1.25-1.27$)
is considered.

In Table \ref{tab.MatElem} the calculated NMEs are presented. The values
of $M_{F,GT,T}$ and $M_\nu$ differ slightly (within 10\%) with those given 
in \cite{vadimp}, which were obtained without consideration of 
the closure approximation. By glancing Table  \ref{tab.MatElem} 
we see that $M_{F\omega,GT\omega,T\omega} \simeq M_{F,GT,T}$ and $M_{\nu\omega}\simeq M_\nu$
as for the average neutrino momentum q = 100 MeV and used average energy 
of intermediate nuclear states we have
${q}/{\left(q +\bar{E}_n -(E_i+E_f)/2\right)} \simeq 1$. The absolute value
of $M_{Fq,GTq,Tq}$ is smaller in comparison with $M_{F,GT,T}$ by about 50\%
for Fermi NMEs and by about factor two in the case of Gamow-Teller
and tensor NMEs.  From Table \ref{tab.MatElem} it follows that there is
a significant difference between results of this work and 
the QRPA NMEs of Ref. \cite{muto89}, especially in the case of $^{100}$Mo.
This difference can
be attributed to the progress achieved in the $0\nu\beta\beta$-decay formalism
due to inclusion of higher order terms of nucleon currents \cite{StefA,Simdva},
the way of adjusting the parameters of nuclear Hamiltonian \cite{Vadim03},
description of short-range correlations \cite{src09} and restoration of
the isospin symmetry \cite{vadimp}.

Nuclear matrix elements $M_{2^-}$, $M_{1^+}$ ($\lambda$ mechanism)
and $M_{\nu}$ ($m_{\beta\beta}$ mechanism) for 10 nuclei under consideration
are given in Table \ref{tab.MatElem} and displayed in Fig. \ref{fig:nme}.
We note a rather good agreement between $M_{2^-}$ and $M_{\nu}$ for all calculated
nuclear systems. It is because the contribution of $M_{1^+}$ to $M_{2^-}$
is suppressed by factor 9 and as a result $M_{2^-}$ is governed by 
the $M_{\nu\omega}$ contribution (see Eq. (\ref{nulamnme})). Values of 
$M_{1^+}$ exhibit similar systematic behavior in respect to considered nuclei 
as values of  $M_{\nu}$ and $M_{2^-}$, but they are suppressed by about factor 2-3
(with exception of $^{48}$Ca).

The importance of the $m_{\beta\beta}$ and $\lambda$ mechanisms depends, respectively,
not only on values of $\eta_\nu$ and $\eta_\lambda$ parameters, which are unknown,
but also on values of coefficients $C_I$
(I=$mm$, $m\lambda$, $\lambda\lambda$), which are listed 
for all studied nuclei in Table \ref{tab.MatElem}. They have been obtained by using 
improved values of phase-space factors $G_{0k}$ (k=1, 2, 10 and 11)  
from \cite{StefA}. We note that the squared value of 
$M_{GT}$ and fourth power of axial-vector coupling constant $g_A$ 
are included in the definition of coefficient $C_I$ unlike in \cite{StefA}.
We see that $C_{\lambda\lambda}$ is always larger when compared with $C_{mm}$.
The absolute value of $C_{m\lambda}$ is significantly smaller than $C_{mm}$ and
$C_{\lambda\lambda}$. This fact points out on less important contribution to
the $0\nu\beta\beta$-decay rate from the interference of $m_{\beta\beta}$
and $\lambda$ mechanisms.

For 10 nuclei of experimental interest the decomposition of coefficient
$C_{\lambda\lambda}$ (see Eq. (\ref{halflifem})) on partial contributions $C^{0k}_I$
associated with phase-space factors $G_{0k}$ (k=2, 10 and 11) is shown in
Fig. \ref{fig:Ccomp}. By glancing the plotted ratio $C^{0k}_I/C_I$ we see 
that $C_{\lambda\lambda}$ is dominated by a single contribution associated with
the phase-space factor $G_{02}$. From this and above analysis it follows that
$0\nu\beta\beta$-decay half-life to a good accuracy can be written as
\bee \label{halflifem2}
\left[T_{1/2}^{0\nu}\right]^{-1} &=& 
\left( \eta_\nu^2 ~+~ \eta_\lambda^2 f_{\lambda m}\right)~ C_{mm} \nonumber\\
&\simeq& \left( \eta_\nu^2 ~+~ \eta_\lambda^2~ f^G_{\lambda m}\right)
~g_A^4 ~M_\nu^2 ~G_{01}
\label{factml}
\eee
with
\begin{equation}
  f_{\lambda m} = \frac{C_{\lambda\lambda}}{C_{mm}} \simeq f^G_{\lambda m}= \frac{G_{02}}{G_{01}}.
\end{equation}

\begin{center}
  \begin{figure}[!t]
  \includegraphics[width=8cm]{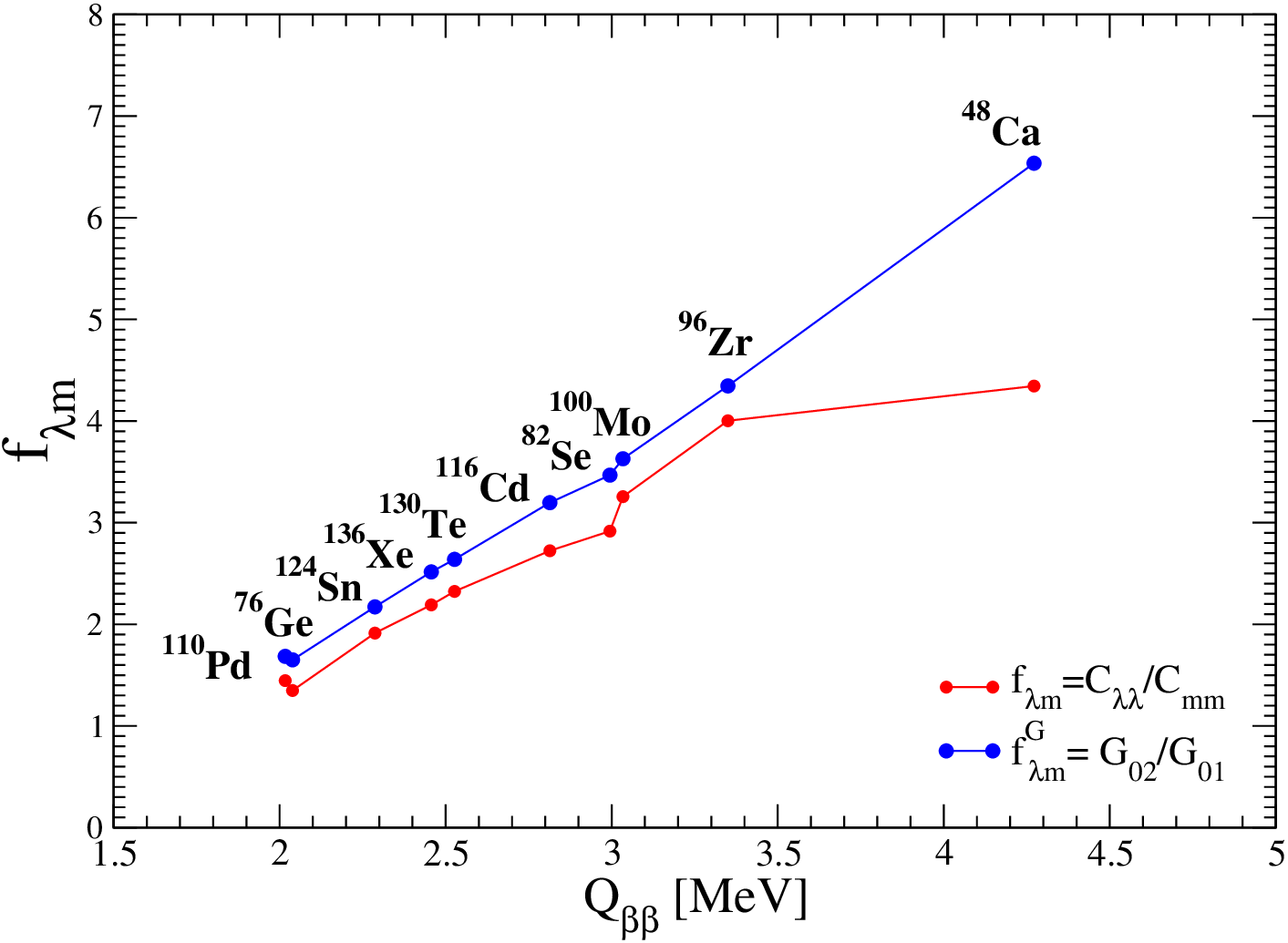}    
\caption{(Color online) The factor $f_{\lambda m}$ (see Eq. (\ref{factml})) as function of Q-value
      of the double beta decay process ($Q_{\beta\beta}$) plotted from the numbers of Table
      \ref{tab.MatElem}. 
  }
\label{fig:fm}
  \end{figure}
\end{center}

For a given isotope the factor $f_{\lambda m}$ reflects relative sensitivity to the $m_{\beta\beta}$
and $\lambda$ mechanisms and $f^G_{\lambda m}$ is its approximation, which does not depend on NMEs.
The values  $f_{\lambda m}$ and $f^G_{\lambda m}$ are tabulated
in Table \ref{tab.MatElem} and plotted as function of $Q_{\beta\beta}$ in Fig. \ref{fig:fm}. 
We see that $f_{\lambda m}$ depends only weakly on involved nuclear matrix elements
(apart for the case of $^{48}$Ca)  what follows from a comparison of $f_{\lambda m}$
with $f^G_{\lambda m}$. The value of $f_{\lambda m}$ is mainly determined by the Q-value
of double beta decay process. From 10 analyzed nuclei the largest value of $f_{m\lambda}$
is found for $^{48}$Ca and the smallest value for $^{76}$Ge. A larger value of $f_{\lambda m}$ means
increased sensitivity to $m_{\beta\beta}$ mechanism in comparison to $\lambda$ mechanism
and vice versa.

\begin{widetext}

\begin{table}[!t]
  \begin{center}
    \caption{
      Upper bounds on the effective Majorana neutrino mass $m_{\beta\beta}$
      and parameter $\eta_\lambda$ associated with right-handed
      currents mechanism  imposed by current constraints on the $0\nu\beta\beta$-decay
      half-life for nuclei of experimental interest. The calculation is performed with
      NMEs obtained within the QRPA with partial restoration of the isospin symmetry
      (see Table \ref{tab.MatElem}).
      The upper limits on $m_{\beta\beta}$ and $\eta_\lambda$ are deduced for a coexistence
      of the $m_{\beta\beta}$ and $\lambda$ mechanisms (Maximum) and for the case 
      $\eta_\lambda  = 0$ or $\eta_\nu = 0$ (On axis).     
      $g_A=1.269$ and CP conservation ($\psi=0$) are assumed. 
      \label{tab.t12}}
 \begin{tabular}{lcccccccc}\hline\hline
 & & $^{48}$Ca & $^{76}$Ge & $^{82}$Se & $^{100}$Mo & $^{116}$Cd  & $^{130}$Te &
   $^{136}$Xe   \\ \hline
   $T^{0\nu-exp}_{1/2}$ [yrs]   & &
     $2.0~10^{22}$ &    $5.3~10^{25}$  &  $2.5~10^{23}$  &  $1.1~10^{24}$  &    
   $1.7~10^{23}$ &    $4.0~10^{24}$ &   $1.07~10^{26}$     \\
   Ref.       & & \cite{expCa} &  \cite{expGe}  & \cite{expSe} & \cite{expMo}
   &  \cite{expCd}    &  \cite{expTe}   &   \cite{expXe}       \\            
                    & &     &     &     &     &      &     &         \\         
   $m_{\beta\beta}$  [eV]    & 
   &   23.8 &   0.185  & 1.45 & 0.484    &  1.55    &  0.379   & 0.128          \\                 
 $\eta_\lambda$         & 
   & $2.24~10^{-5}$ & $3.11~10^{-7}$   & $1.65~10^{-6}$  & $5.25~10^{-7}$    &$1.84~10^{-6}$
   & $4.87~10^{-7}$    & $1.70~10^{-7}$        \\
   & &    \multicolumn{7}{c}{for $\eta_\lambda  = 0$ }     \\         
     $m_{\beta\beta}$  [eV]    & &
 23.7 &  0.182   &  1.43   & 0.477  & 1.53   & 0.374    &   0.126     \\              
                    & &    \multicolumn{7}{c}{for  $m_{\beta\beta}  = 0$ }    \\               
  $\eta_\lambda$         & &
 $2.23~10^{-5}$  &  $3.07~10^{-7}$  & $1.63~10^{-6}$  & $5.18~10^{-7}$ &
 $1.81~10^{-6}$ & $4.80~10^{-7}$  &  $1.67~10^{-7}$   \\              
\hline\hline   
      \end{tabular}
  \end{center}
\end{table}

\end{widetext}      

Upper bounds on the effective neutrino mass $m_{\beta\beta}$ and right-handed
current coupling strength $\eta_\lambda$ are deduced from
experimental half-lives of the $0\nu\beta\beta$-decay by using the coefficients
$C_{mm}$, $C_{m\lambda}$ and $C_{\lambda\lambda}$ of Table \ref{tab.MatElem}. The
maximum and the value on axis ($m_{\beta\beta}=0$ or $\eta_\lambda =0$) 
are listed in Table \ref{tab.t12}. The decays of $^{136}$Xe and $^{76}$Ge
set the sharpest limit $m_{\beta\beta}\le$ 0.13 eV and 0.18 eV, and
$\eta_\lambda\le 1.7~10^{-7}$ and $3.1~10^{-7}$, respectively. 
These are more stringent than those deduced from other experimental sources.

\begin{figure}[!t]
  \includegraphics[width=8cm]{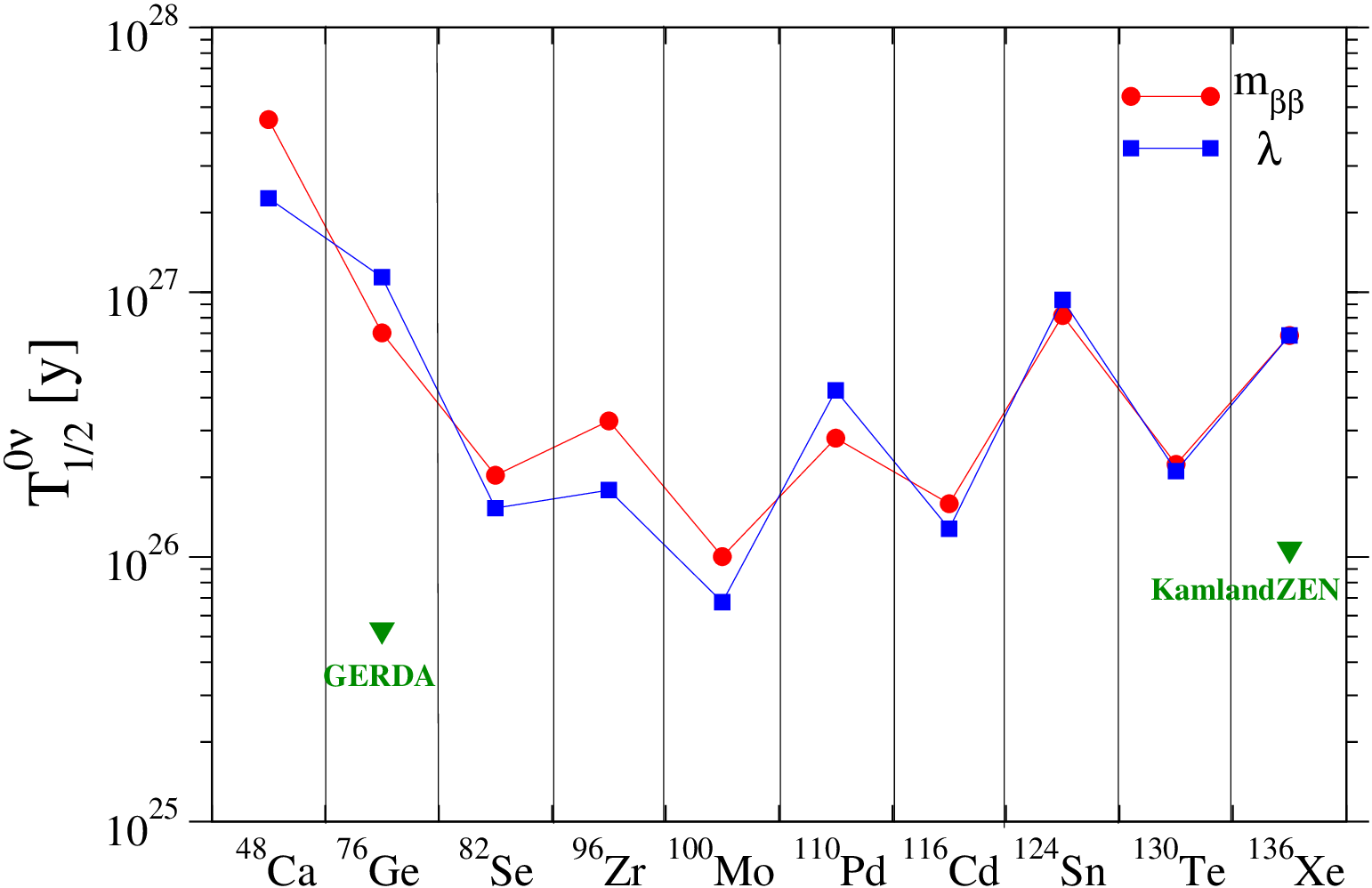}
  \caption{(Color online) The $0\nu\beta\beta$-decay half-lives of nuclei of
    experimental interest calculated for $m_{\beta\beta}$ (red circle) and $\lambda$ (blue square)
    mechanisms by assuming an illustrative case of observation $0\nu\beta\beta$-decay
    of $^{136}$Xe with half-life $T^{0\nu}_{1/2}=6.86~10^{26}$ yrs ($m_{\beta\beta}$ = 50 meV
    or $\eta_\lambda$ = $9.8~10^{-8}$). The current experimental limits on $0\nu\beta\beta$-decay half-life
    of $^{76}$Ge (the GERDA experiment) and  $^{136}$Xe (the Kamland-Zen experiment)  are displayed
    with green triangles. 
  } 
\label{fig.t12}
\end{figure}

It is well known that by measuring different characteristics, namely energy and angular distributions
of two emitted electrons, it is possible to identify which of $m_{\beta\beta}$ and $\lambda$ mechanisms
is responsible for $0\nu\beta\beta$-decay \cite{StefA,Doies}. It might be achieved only by some of future
$0\nu\beta\beta$-decay experiments, e.g. the SuperNEMO \cite{Super} or NEXT \cite{Next}. A relevant
question is whether the underlying $m_{\beta\beta}$ or $\lambda$ mechanism can be
revealed by observation of the $0\nu\beta\beta$-decay in a series of different isotopes.
In Fig. \ref{fig.t12} this issue is addressed by an illustrative case of observation
of the $0\nu\beta\beta$-decay of $^{136}$Xe with half-life $T^{0\nu}_{1/2}=6.86~10^{26}$ yrs,
which can be associated with $m_{\beta\beta}$ = 50 meV or $\eta_\lambda$ = $9.8~10^{-8}$.
The $0\nu\beta\beta$-decay half-life predictions associated with a dominance of $m_{\beta\beta}$
and $\lambda$ mechanisms exhibit significant difference for some nuclear systems. 
We see that by observing, e.g., the $0\nu\beta\beta$-decay of $^{100}$Ge and $^{100}$Mo with sufficient accuracy
and having calculated relevant NMEs with uncertainty below  30\%, it might be  possible to conclude,
whether the $0\nu\beta\beta$-decay is due to $m_{\beta\beta}$ or $\lambda$ mechanism.

Currently, the uncertainty in calculated $0\nu\beta\beta$-decay NMEs can be estimated up to factor
of 2 or 3 depending on the considered isotope as it follows from a comparison
of results of different nuclear structure approaches \cite{ves16}. The improvement of the calculation of
double beta decay NMEs is a very important and challenging problem.
There is a hope that due to a recent progress in nuclear structure theory (e.g., ab initio
methods) and increasing computing power the calculation of the $0\nu\beta\beta$-decay NMEs with
uncertainty of about 30 \%  might be achieved in future.  

%
\section{The lepton number violating parameters within the seesaw and normal hierarchy}
%

The $6\times 6$ unitary neutrino mixing matrix $\mathcal{U}$ (see Eq. (\ref{maticazmies}))
can be parametrized with 15 rotational angles and 10 Dirac and 5 Majorana CP violating phases.
For the purpose of study different LRSM contributions to the $0\nu\beta\beta$-decay the mixing
matrix $\mathcal{U}$ is usually decomposed as follows \cite{Xing}
\bee
 {\mathcal{U}} &=& \left(
\begin{array}{ll}
\mathbf{1} & \mathbf{0}\\
 \mathbf{0} & U_0\\
 \end{array}
\right)
\left(
\begin{array}{ll}
A & R\\
S & B\\
 \end{array}
\right)
\left(
\begin{array}{ll}
V_0 & \mathbf{0}\\
 \mathbf{0} & \mathbf{1}\\
 \end{array}
\right)
\label{maticaXing}.
\eee
Here, $\mathbf{0}$ and $\mathbf{1}$ are the $3\times 3$ zero and identity matrices, respectively. 
The parametrization of matrices A, B, R and S and corresponding orthogonality relations are given
in \cite{Xing}.

If A = $\mathbf{1}$, B =$\mathbf{1}$, R = $\mathbf{0}$ and S = $\mathbf{0}$, there would be
a separate mixing of light and heavy neutrinos, which would participate only in left and right-handed
currents, respectively. In this case we get $\eta_\lambda = 0$, i.e., the $\lambda$
mechanism is forbidden. 

\begin{figure}[!t]
  \includegraphics[width=8cm]{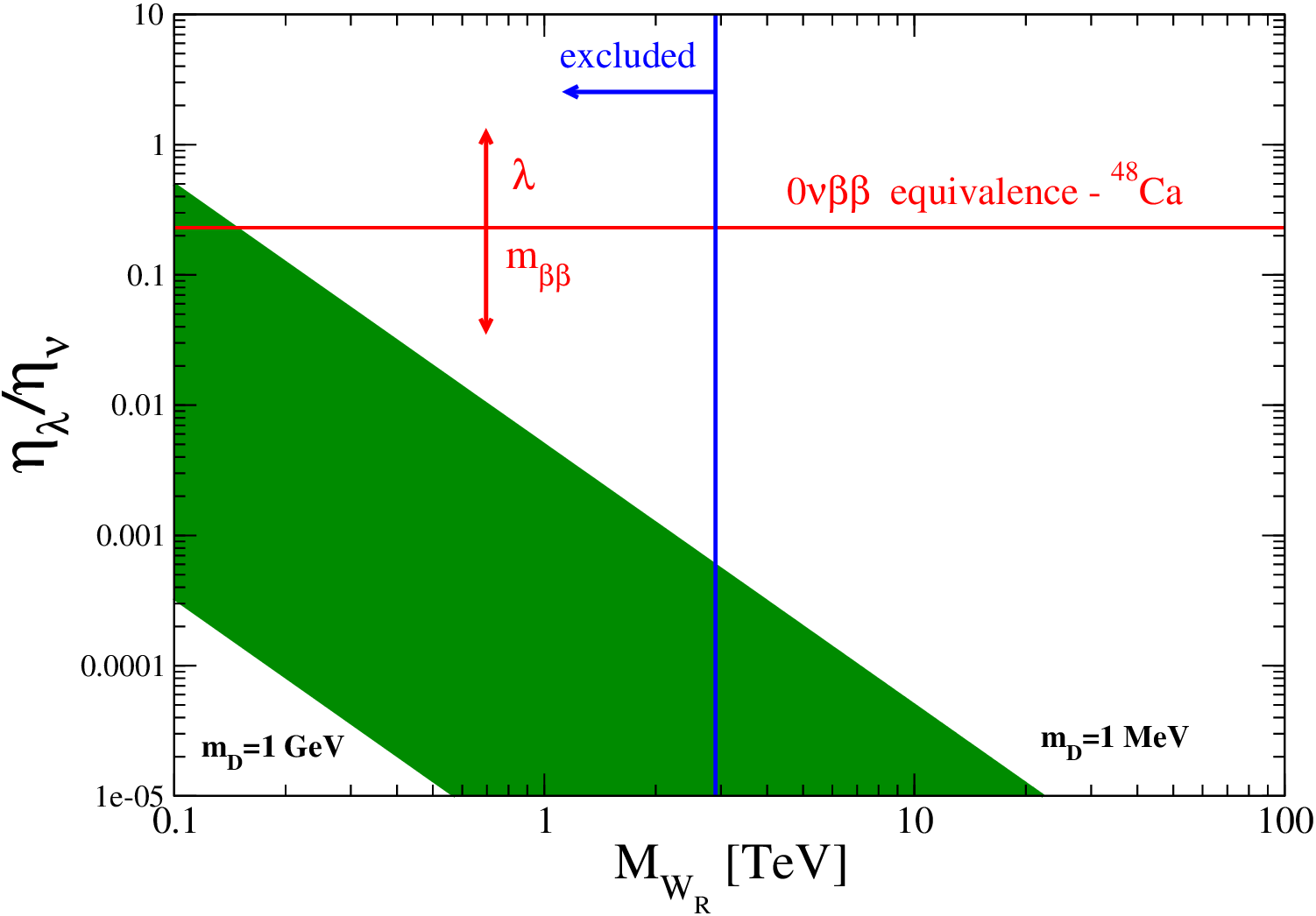}  
  \caption{
    (Color online) The allowed range of values for the ratio  $\eta_\lambda /{\eta_{\nu}}$ (in green)
  as a function of the mass of the heavy vector boson $M_{W_R}$. The line of the $0\nu\beta\beta$ equivalence
  corresponds to the case of equal importance of both  $m_{\beta\beta}$ and $\lambda$ mechanisms
  in the $0\nu\beta\beta$-decay rate.
  } 
  \label{fig.lnv}
\end{figure}

If masses of heavy neutrinos are above the TeV scale, the mixing angles responsible for mixing of light
and heavy neutrinos are small. By neglecting the mixing between different generations of light and heavy neutrinos,
the unitary mixing matrix $\mathcal{U}$ takes the form
\bee
\mathcal{U}&=& \left(
\begin{array}{cc}
U_0 & \frac{m_D}{m_{LNV}}~ \mathbf{1} \\
 - \frac{m_D}{m_{LNV}}~\mathbf{1} & V_0 \\
 \end{array}
\right). 
\label{specmixing}
\eee
Here, $m_D$ represents energy scale of charged leptons and $m_{LNV}$ is the total lepton number violating scale,
which corresponds to masses of heavy neutrinos. We see that $U=U_0$ can be identified to a good approximation
with the PMNS matrix and $V_0$ is its analogue for heavy neutrino sector. Due to unitarity condition we find
$V_0 = U^\dagger_0$. Within this scenario of neutrino mixing the
effective lepton number violating parameters $\eta_\nu$ ($m_{\beta\beta}$
mechanism) and $\eta_\lambda$ ($\lambda$ mechanism)  are given by
\begin{eqnarray}
  \eta_{\nu}  
  &=& \frac{m_D}{m_e} \frac{m_D}{m_{LNV}} ~\zeta_m, \nonumber\\
  \eta_{\lambda}
  &=& \left(\frac{M_{W_L}}{M_{W_R}}\right)^2 \frac{m_D}{m_{LNV}}  \zeta_\lambda
\end{eqnarray}  
with 
\begin{eqnarray}
  \zeta_m &=& \left|\sum_{j=1}^3 U_{ej}^2 \frac{m_j m_{LNV}}{m_D^2}\right|,\nonumber\\
  \zeta_\lambda &=& |\sum_{j=1}^3 U_{ej}| = 0.14-1.5.
\label{zeta}
\end{eqnarray}

The importance of $m_{\beta\beta}$ or $\lambda$-mechanism can be judged from the ratio
\begin{eqnarray}
  \frac{\eta_{\lambda}}{\eta_{\nu}} = \left(\frac{M_{W_L}}{M_{W_R}}\right)^2 \frac{m_e}{m_D}
  \frac{\zeta_\lambda}{\zeta_m}.   
\label{approx}
\end{eqnarray}
It is naturally to assume that $\zeta_m \approx 1$ and to consider the upper bound for the factor
$\zeta_\lambda$, i.e., there is no anomaly cancellation among terms, which constitute these
factors. Within this approximation $\eta_\lambda /{\eta_{\nu}}$ does not depend on scale of
the lepton number violation $m_{LNV}$ and is plotted in Fig. \ref{fig.lnv}.
The Dirac mass $m_D$ is assumed to  be within the range $1~ \textrm{MeV} < m_D < 1~ \textrm{GeV}$.
The flavor and CP-violating processes of kaons and B-mesons make it possible to deduce lower bound
on the mass of the heavy vector boson $M_{W_2} > 2.9$ TeV \cite{Dev}. From Fig. \ref{fig.lnv}
it follows that within accepted assumptions the $\lambda$ mechanism is  practically excluded
as the dominant mechanism of the $0\nu\beta\beta$-decay. 

In this section the light-heavy neutrino mixing of the strength $m_D/m_{LNV}$ is considered. 
However, we note that there are models with heavy neutrinos mixings where strength of the mixing
decouples from neutrino masses \cite{pilaf92,gluza02,deppi11,vissa12,dev13}. This subject goes
beyond the scope of this paper. 

%
\section{Summary and Conclusions}
%

The left-right symmetric model of weak interaction is an attractive extension of the Standard Model, which
may manifest itself in the TeV scale. In such case the Large Hadron Collider can determine the
right-handed neutrino mixings and heavy neutrino masses of the seesaw model. The LRSM 
predicts new physics contributions to the $0\nu\beta\beta$ half-life due to exchange of light and
heavy neutrinos, which can be sizable.

In this work the attention was paid to the $\lambda$ mechanism of the $0\nu\beta\beta$-decay, which
involves left-right neutrino mixing through mediation of light neutrinos.  The recently improved
formalism of the $0\nu\beta\beta$-decay concerning this mechanism was considered.
For 10 nuclei of experimental interest NMEs were calculated within the QRPA with a partial restoration
of the isospin symmetry. It was found that matrix elements governing
the conventional $m_{\beta\beta}$ and $\lambda$ mechanisms are comparable and that the $\lambda$
contribution to the decay rate can be associated with a single phase-space factor. A simplified
formula for the $0\nu\beta\beta$-decay half-life is presented (see Eq. \ref{factml})), which neglects
the suppressed contribution from the interference of both mechanisms. In this expression
the $\lambda$ contribution to decay rate is weighted by the factor $f_{\lambda m}$, which reflects relative
sensitivity to the $m_{\beta\beta}$ and $\lambda$ mechanisms for a given isotope and
depends only weakly on nuclear physics input. It is manifested that measurements of
$0\nu\beta\beta$-decay half-life on multiple isotopes with largest deviation in the factor
$f_{\lambda m}$ might allow to distinguish both considered mechanisms, if involved NMEs are known
with sufficient accuracy.  

Further, upper bounds on effective lepton number violating parameters $m_{\beta\beta}$ ($\eta_\nu$)
and $\eta_\lambda$ were deduced from current lower limits on experimental half-lives of
the $0\nu\beta\beta$-decay. The ratio $\eta_\lambda/\eta_\nu$ was studied as function of
the mass of heavy vector boson $M_{W_R}$ assuming that there is no mixing among
different generations of light and heavy neutrinos. It was found that if 
the value of Dirac mass $m_D$ is within the range $1~ \textrm{MeV} < m_D < 1~ \textrm{GeV}$,
the current constraint on $M_{W_R}$ excludes the dominance of the $\lambda$ mechanism
in the $0\nu\beta\beta$-decay rate for the assumed neutrino mixing scenario.
 
\section*{Funding}
This work is supported by the VEGA Grant Agency of the Slovak Republic un- der Contract No. 1/0922/16,
by Slovak Research and Development Agency under Contract No. APVV-14-0524, RFBR Grant
No. 16-02-01104, Underground laboratory LSM - Czech participation to European-level research
infrastructue CZ.02.1.01/0.0/0.0/16 013/0001733.

%

%

\end{document}